\documentclass[aps,pra,twocolumn,superscriptaddress]{revtex4-1}

\usepackage{bm}
\usepackage{mathrsfs}
\usepackage{amsmath}
\usepackage{color}
\usepackage{amssymb}
\usepackage[colorlinks]{hyperref}
\usepackage{pdfpages}
\usepackage{dsfont}
\usepackage[colorlinks]{hyperref}

\newcommand{\ket}[1]{|#1\rangle}

\begin{document}
\title{Entangling two transportable neutral atoms via local spin exchange}

\author{A. M.~Kaufman}
\affiliation{JILA, National Institute of Standards and Technology and University of Colorado}
\affiliation{Department of Physics, University of Colorado, Boulder, Colorado 80309, USA}
\author{B. J.~Lester}
\affiliation{JILA, National Institute of Standards and Technology and University of Colorado}
\affiliation{Department of Physics, University of Colorado, Boulder, Colorado 80309, USA}
\author{M. Foss-Feig}
\affiliation{Joint Quantum Institute and the National Institute of Standards and Technology, Gaithersburg, Maryland, 20899, USA}
\author{M. L. Wall}
\affiliation{JILA, National Institute of Standards and Technology and University of Colorado}
\affiliation{Department of Physics, University of Colorado, Boulder, Colorado 80309, USA}
\author{A. M. Rey}
\affiliation{JILA, National Institute of Standards and Technology and University of Colorado}
\affiliation{Department of Physics, University of Colorado, Boulder, Colorado 80309, USA}
\author{C. A. Regal}
\affiliation{JILA, National Institute of Standards and Technology and University of Colorado}
\affiliation{Department of Physics, University of Colorado, Boulder, Colorado 80309, USA}


\date{\today}




\maketitle 

\begin{figure*}
	\centering
	\includegraphics[scale=0.4]{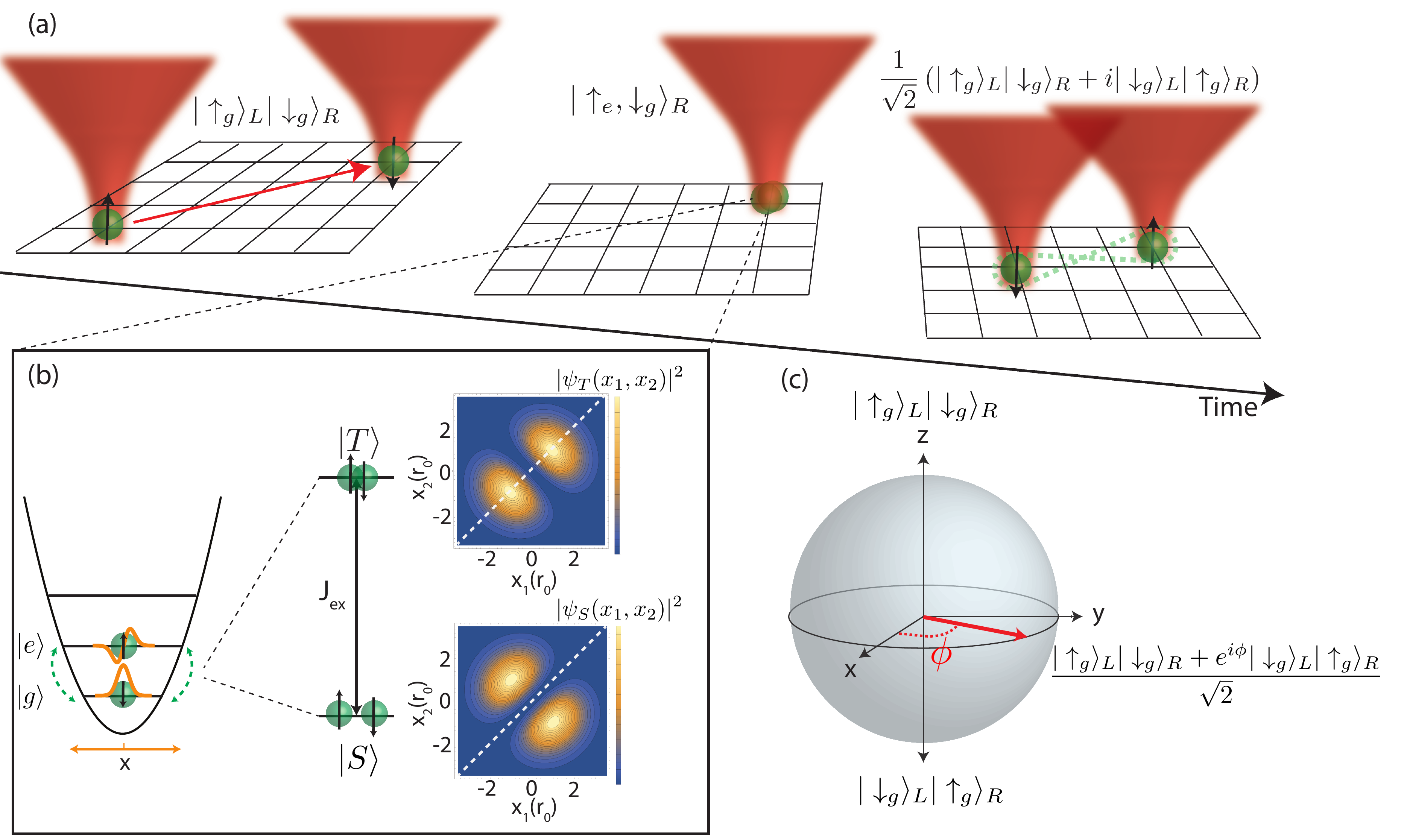}
	\caption{Experimental overview. (a) Two particles, trapped in separate optical tweezers, are initiated in their motional ground state and in opposing spin states (left). Dynamical reconfiguration of optical tweezer traps merges the particles into the same optical tweezer in specific motional states, and spin exchange entangles the particles. The atoms are separated into different optical tweezers, and the entanglement between the particles is experimentally detected (right).  (b) Exchange interaction. Two atoms of opposite spin are prepared in the ground (g) and excited (e) motional states of the $x$ direction. For bosonic atoms the singlet channel anti-symmetrizes in the motional degree of freedom, yielding $\psi_S ({\bf r_1}, {\bf r_2}) = (\psi_g ({\bf r_1})\psi_e ({\bf r_2})-\psi_e ({\bf r_1})\psi_g ({\bf r_2}))/\sqrt{2}$. From one-dimensional spatial wave functions we see that the anti-symmetrization prevents the two particles at positions $x_1$ and $x_2$ (in units of the oscillator length $r_0$) from occupying the same position (dashed white line), which is true for all choices of $y,z$. This leads to zero interaction energy for the singlet channel and $J_{\rm ex}$ for the triplet channel with associated wave function $\psi_T({\bf r_1}, {\bf r_2}) = (\psi_g ({\bf r_1})\psi_e ({\bf r_2})+\psi_e ({\bf r_1})\psi_g ({\bf r_2}))/\sqrt{2}$.  (c) Studying entanglement of the separated particles. On the Bloch sphere, the exchange interactions ideally result in a state pointing along the $\pm~y$-axis, but in general can point in any direction in the equatorial plane with associated coherence angle $\phi$.}
	\label{fig:exp}
\end{figure*} 

To advance quantum information science a constant pursuit is the search for physical systems that meet the stringent requirements for creating and preserving quantum entanglement.  In atomic physics, robust two-qubit entanglement is typically achieved by strong, long-range interactions in the form of Coulomb interactions between ions or dipolar interactions between Rydberg atoms~\cite{Sackett2000, Liebfried2003a, Wilk2010, Isenhower2010}. While these interactions allow fast gates, atoms subject to these interactions must overcome the associated coupling to the environment and cross-talk among qubits~\cite{Monroe1995,Blakestad2009,home2009,Beguin2013}. Local interactions, such as those requiring significant wavefunction overlap, can alleviate these detrimental effects yet present a new challenge: To distribute entanglement, qubits must be transported, merged for interaction, and then isolated for storage and subsequent operations.  Here we show how, via a mobile optical tweezer, it is possible to prepare and locally entangle two ultracold neutral atoms, and then separate them while preserving their entanglement~\cite{Hayes2007,Anderlini2007,Weitenberg2011a}. While ground-state neutral atom experiments have measured dynamics consistent with spin entanglement~\cite{Mandel2003a,Anderlini2007, Fukuhara2013}, and detected entanglement with macroscopic observables~\cite{Lucke2011,Strobel2014}, we are now able to demonstrate position-resolved two-particle coherence via application of a local gradient and parity measurements~\cite{Sackett2000}; this new entanglement-verification protocol could be applied to arbitrary spin-entangled states of spatially-separated atoms~\cite{Endres,Fukuhara2015}. The local entangling operation is achieved via ultracold spin-exchange interactions~\cite{Hayes2007,Anderlini2007,Weitenberg2011a}, and quantum tunneling is used to combine and separate atoms. Our toolset provides a framework for dynamically entangling remote qubits via local operations within a large-scale quantum register.

Internal spin states of particles provide a robust and long-lived storage for quantum information.  While engineered spin-dependent interactions can realize entangling gates between spins, they also predispose the system to strong environmental coupling and decoherence~\cite{Hayes2007}.  Spin-exchange interactions, which arise from a combination of quantum statistics and spin-independent forces, afford a promising alternate route to entanglement generation, and have been explored with both electrons in quantum dots and atoms ~\cite{Divincenzo2000,petta2005, Hayes2007, Anderlini2007,Trotzky2008,Weitenberg2011a}. When two particles interact, their interaction energy depends on the spatial symmetry of the two-particle wave function. If the particles have spin but are otherwise identical, the symmetry of the two-particle spin state directly determines the spatial symmetry of their wave function:  For repulsive interactions, two bosons (fermions) of opposite spin in a triplet configuration experience enhanced (suppressed) interactions, while the converse occurs for the singlet spin state (Figure~\ref{fig:exp}b).  By preparing a superposition of the triplet and singlet, dynamical quantum beats result in the exchange of spin between the particles.  In our experiment we prepare atoms of opposing spin in the lowest two motional states ($e$ and $g$) of an optical tweezer potential that we represent as $\ket{\!\uparrow_e ,\!\downarrow_g}$, which results in equal population of the singlet and triplet spatial wave functions $\psi_S ({\bf r_1}, {\bf r_2})$ and $\psi_T ({\bf r_1}, {\bf r_2})$. The difference in the contact interaction energy between these  states yields spin-exchange dynamics at a rate $J_{\rm ex}$, which depends on the $s$-wave scattering length and two-particle density (see supplementary materials). Specifically, it gives rise to the dynamics
\begin{equation} 
 \vert \psi(t) \rangle = \ket{\!\uparrow_e ,\!\downarrow_g} \cos(J_{\rm ex} t/2\hbar) +i\ket{\!\downarrow_e, \!\uparrow_g}\sin(J_{\rm ex}t/2\hbar).
\end{equation}
This evolution is associated with the effective spin-dependent Hamiltonian $H= J_{\mathrm{ex}} {\bf S_{e}}\cdot {\bf S_{g}}$, where ${\bf S_{e}}$ and ${\bf S_{g}}$ are the spin-operators for the respective motional states. A spin-entangled state $(\ket{\!\uparrow_e,\!\downarrow_g} +i\ket{\!\downarrow_e ,\!\uparrow_g})/\sqrt{2}$ can be created by allowing the state to evolve for an exchange time of $\pi\hbar/2J_{\rm ex}$. 

We schematically represent the experiment in Figure~\ref{fig:exp}a, in which separated optical tweezers on the left (L) and right (R) each containing a single atom are dynamically reconfigured to produce spin-exchange dynamics and non-local entanglement. Atoms are combined in the right optical tweezer where spin exchange creates the entangled state $(\ket{\!\uparrow_e,\!\downarrow_g}_R +i\ket{\!\downarrow_e ,\!\uparrow_g}_R)/\sqrt{2}$.  Importantly, in our experiments we convert this entanglement into spatial-spin correlations by separating the atoms into two tweezers (Figure~\ref{fig:exp}a) to yield a state $(\ket{\!\uparrow_g}_L \ket{\!\downarrow_g}_R +i\ket{\!\downarrow_g}_L\ket{\!\uparrow_g}_R)/\sqrt{2}$.  While verification of such entanglement is a standard tool in ion and Rydberg experiments~\cite{Sackett2000, Wilk2010, Isenhower2010,Kotler2014}, spatially-resolved detection of the entanglement present in interacting systems of ground-state neutral atoms is challenging. Theoretical proposals have studied ways of detecting spatial~\cite{Daley} and spin~\cite{Endres} entanglement, and very recently experimental progress has been made using a quantum gas microscope~\cite{Fukuhara2015}. We devised a protocol that yields rotations of a two-qubit entangled state on the associated Bloch sphere via a combination of a magnetic-field gradient and global spin rotations (Figure~\ref{fig:exp}c), allowing detection of basis-independent correlations for arbitrary entangled states of the form $\frac{1}{\sqrt{2}}\left (\ket{\!\uparrow_g}_L\ket{\!\downarrow_g}_R+e^{i\phi}\ket{\!\downarrow_g}_L\ket{\!\uparrow_g}_R \right )$. Our protocol is applicable to qubit pairs in a large quantum register~\cite{Weitenberg2011a}, to interacting spins in a Bose-Hubbard chain~\cite{Endres, Fukuhara2013}, and to strongly interacting fermions featuring anti-ferromagnetic correlations~\cite{Greif2013, Jochim2015, Hulet2015}. 

\begin{figure*}
	\centering
	\includegraphics[scale=0.75]{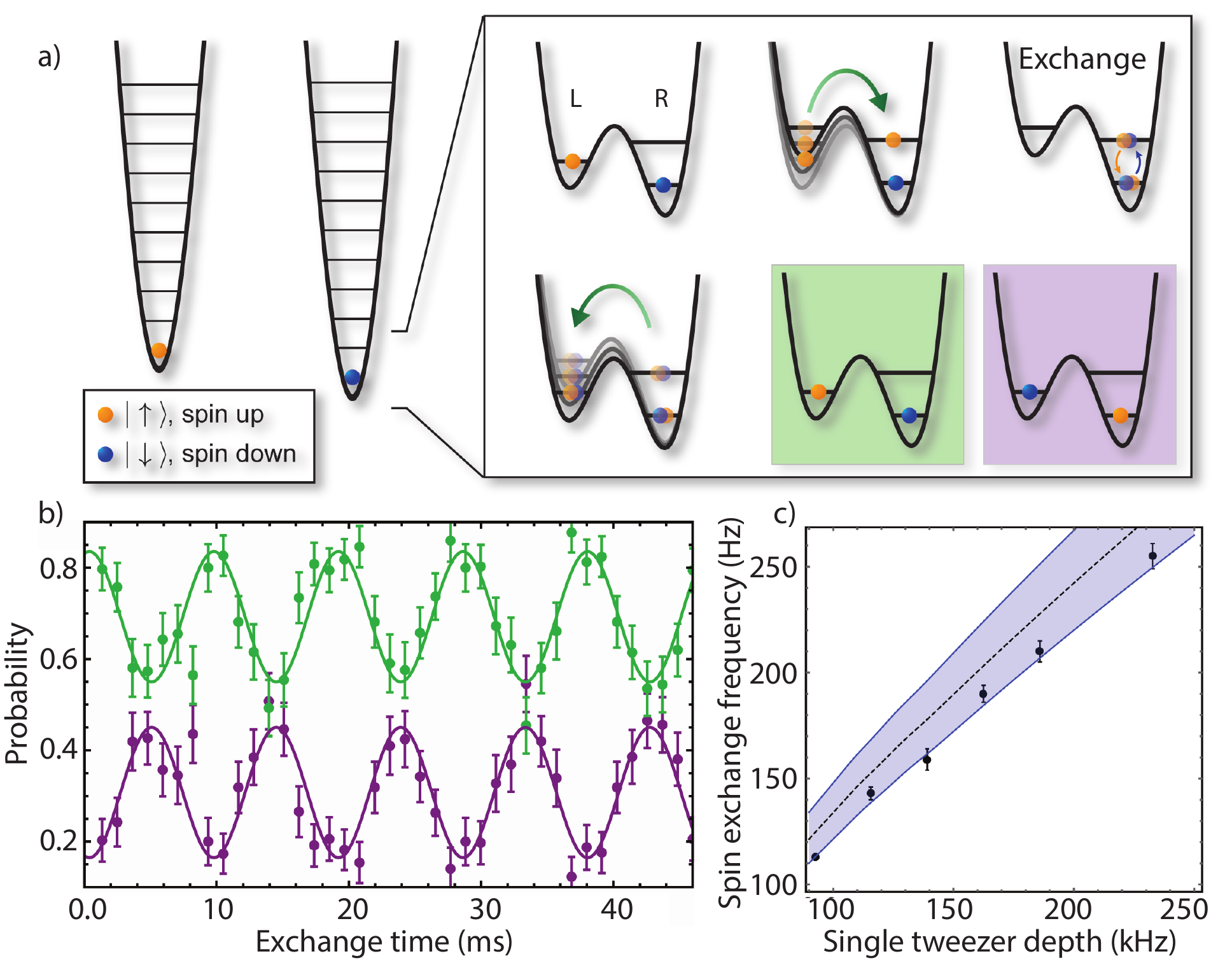}
	\caption{Direct observation of spin-exchange dynamics between two atoms. (a) Preparation of motional state configuration and detection in the double-well potential. (b) Using post-selection on our spin preparation, we plot the likelihood to measure the state $\ket{\!\!\uparrow_g}_L \ket{\!\!\downarrow_g}_R$ (green) and $\ket{\!\!\downarrow_g}_L \ket{\!\!\uparrow_g}_R$ (purple) as a function of time between the end of the first adiabatic passage and the beginning of the second.  (c) Measured spin-exchange oscillation frequency as a function of optical tweezer depth. The dashed black line is the predicted exchange frequency from a parameter-free model of the potential (see supplementary materials); the shaded region indicates the effect of systematic uncertainties on this prediction. All error bars are the standard error in the measurement.}
	\label{fig:exchange}
\end{figure*}

The experiment begins by loading two thermal $^{87}$Rb atoms into two separate optical tweezer potentials~\cite{Schlosser2001}. Each atom is then separately laser cooled to the 3D ground state via Raman-sideband cooling, leaving a 3D ground state fraction of $90(7)\%$~\cite{Monroe1995, Kaufman2012, Kaufman2014}.  The atoms are initialized in opposite spin states with a fidelity in the range of $80-91\%$ via single-atom addressing (see supplementary materials).   We then prepare the pure two-particle state $\ket{\!\uparrow_e,\!\downarrow_g}_R$ on a single tweezer by reducing the separation between the two optical tweezers, and using our capability to control the tunnel coupling between wavelength-scale optical tweezer traps~\cite{Kaufman2014,Jochim2015}. As illustrated in Figure~\ref{fig:exchange}a, the spacing between the two tweezers is decreased until tunneling is appreciable; however, unlike our previous work~\cite{Kaufman2014}, we dynamically shift to an asymmetric configuration such that the ground state of one optical tweezer is near resonant with the first (radial) excited state of the other optical tweezer. We then perform adiabatic passage by slowly tuning the relative tweezer depths linearly in time,  which coherently transfers the left atom into the right optical tweezer over the 12 ms duration of the ramp (see supplementary materials). 
 
After a desired evolution time in the presence of exchange dynamics, the adiabatic passage is applied in reverse, yielding a motional state mapping of the excited-state atom back into the ground state of the left well. We can then read out the spin populations to verify the exchange oscillations by ejecting atoms in the $\vert \!\!\uparrow \rangle$ state from the optical tweezers, and imaging the atom occupancy in each of the two tweezers. With this procedure, we can ascertain what the spin and motional degrees of freedom of each of the two atoms were when they occupied the same optical tweezer. 

Exchange oscillations in our experiment are shown in Figure~\ref{fig:exchange}b, and show the expected anti-correlated behavior.  For these data, we remove experiments in which imperfections in our spin preparation lead to the spins remaining aligned; such events yield a static contribution to the signal. We observe undamped oscillations out to times as long as $100~\mathrm{ms}$, despite the fact that our single-particle coherence time between $\vert \!\! \uparrow \rangle$ and $\vert \!\!\downarrow \rangle$ is less than $1~\mathrm{ms}$ due to magnetic-field fluctuations. This is an expected feature of the entangled states created by the exchange interaction: homogeneous magnetic-field fluctuations induce a global phase on the two-particle superposition and, as such, leave quantum measurements unaffected.  Hence, the state occupies a so-called ``decoherence free subspace"~\cite{Kielpinski2001,Anderlini2007}. We can also control the frequency of the spin oscillations by modifying the depth of the optical tweezer in which the exchange occurs, which tunes the two-particle density. To study this dependence, we prepare $\ket{\!\uparrow_e,\!\downarrow_g}_R$ and linearly increase the depth of the tweezer in $5~\mathrm{ms}$, allow evolution of the exchange, and then ramp back in reverse and perform the second adiabatic passage. We can model the three-dimensional non-separable potential of our optical tweezer trap~\cite{Wall2015}, and find agreement (Figure~\ref{fig:exchange}c) between the calculated and measured spin-exchange frequency $J_{\rm ex}/(2\pi\hbar)$ (see supplementary materials). 

\begin{figure*}
	\centering
	\includegraphics[scale=0.7]{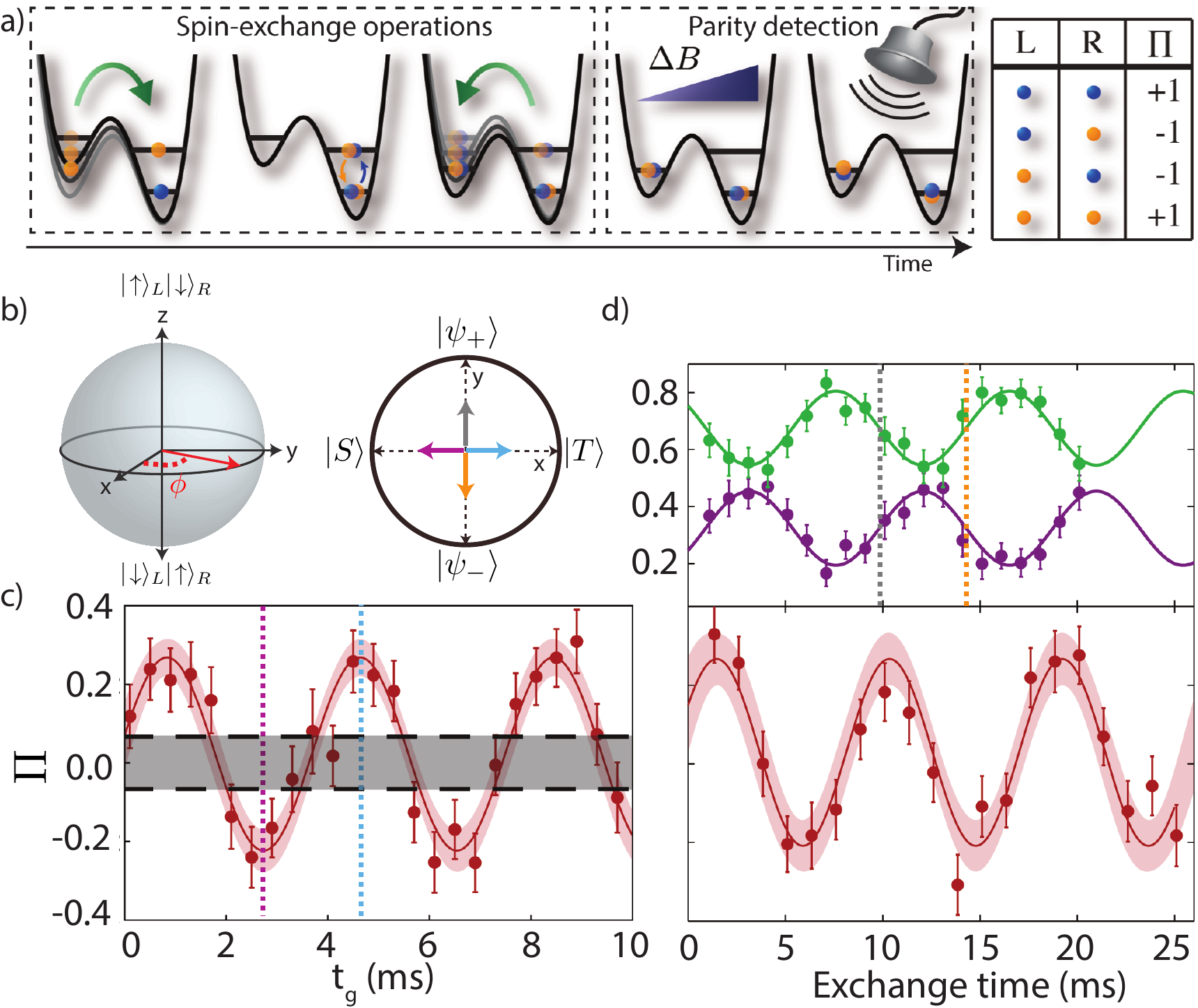}
	\caption{Detection of non-local entanglement. (a) Procedure for creating and detecting entanglement. (b) Highlight of four different Bloch vector orientations in the experiment. The gray and orange orientations correspond to the outcome of the spin-exchange dynamics (after the second adiabatic passage), which yield rotations in the $y$-$z$ plane of the Bloch sphere. The pink and blue orientations correspond to the points of peak parity, and are accessed by applying a magnetic field gradient that rotates the states about the $z$-axis. (c) After creating $\ket{\psi_+}$, we plot the measured parity as a function of gradient time $t_g$. The gray bar is the bound on parity oscillation contrast delineating separable and entangled states, which accounts for imperfect spin-preparation. The colored dashed lines are at times when the associated states indicated in (b) are created. (d) Measured parity as the exchange time is varied and correspondingly the atoms are entangled and unentangled. In the lower plot, we set $t_g$ such that it rotates $\ket{\psi_+}$ to  $\ket{T}$ and then measure the parity. The upper plot is the same experiment without the parity detection, i.e. the protocol of Figure~\ref{fig:exchange}. The dashed lines indicate times when the corresponding states indicated in (b) are produced. The error bars in the data plots are the standard error, and the pink swaths show the 95\% confidence bands.}
	\label{fig:Parity}
\end{figure*} 

The measurements presented thus far have shown correlations in single-particle spin states. However, to ensure that future operations can retain and propagate quantum information, one must verify that the phase coherence within the entangled state is preserved upon separating the particles. The entanglement verification protocol for separated atoms is summarized in Figure~\ref{fig:Parity}a. For explanatory purposes, we first focus on the case when the particles are separated after an exchange time of $t_{\rm ent} = n\pi\hbar/2J_{\rm ex}$ where $n$ is an odd integer.  The entangled state after the second adiabatic passage is $\vert \psi_{\pm} \rangle~=~\frac{1}{\sqrt{2}}(\vert \!\uparrow \rangle_L \vert \!\downarrow \rangle_R \pm i\vert \!\downarrow \rangle_L \vert \!\uparrow \rangle_R)$, omitting from now on the ground-state (g) motional subscripts to simplify notation. The $\vert \psi_{\pm}\rangle$ states correspond to the gray and orange Bloch vectors, respectively, in Figure~\ref{fig:Parity}b.  We then apply a magnetic-field gradient that imposes a difference, $\delta\hbar$, in the $\vert \!\uparrow \rangle\leftrightarrow\vert \!\downarrow \rangle$ single-atom-transition energy between the left and right optical tweezer. By applying the gradient for a time $t_{g}$, a transformation $\vert\psi_{\pm}\rangle \rightarrow \frac{1}{\sqrt{2}} ( \vert \!\uparrow \rangle_L \vert \!\downarrow \rangle_R  \pm ie^{i\delta t_g}\vert \!\downarrow \rangle_L \vert \uparrow \rangle_R)$ is achieved. As a function of $t_g$, the state rotates between the singlet (pink in Figure~\ref{fig:Parity}b) and triplet (blue) with frequency $\delta$. We then apply a global $\pi/2$ pulse in the $\{\vert \!\uparrow \rangle, \vert \! \downarrow \rangle\}$ sub-space.  This pulse maps the singlet back to itself, while it maps the triplet to a Bell state $\frac{i}{\sqrt{2}}(\vert\! \uparrow \rangle_L \vert \!\uparrow \rangle_R  + \vert \!\downarrow \rangle_L \vert \!\downarrow \rangle_R)$. Therefore, by measuring the probability that the spins are aligned or anti-aligned as a function of $t_g$, we can observe singlet-triplet oscillations whose amplitude characterizes the two-particle coherence. We quantify this probability with the parity $\Pi(t_g) = \sum_j P_j(-1)^j$, where $P_j$ is the likelihood to measure $j$ atoms in the spin-down state~\cite{Sackett2000, Liebfried2003a, Wilk2010, Isenhower2010}. The parity is equivalently the projection of the Bloch vector in Figure~\ref{fig:Parity}b onto the $x$-axis prior to the $\pi/2$-pulse, and hence the gradient is essential because, though entangled, the states $\ket{\psi_{\pm}}$ (gray, orange) exhibit zero parity after application of a $\pi/2$-pulse. 

We demonstrate the outcome of the verification protocol on the state $\ket{\psi_+}$ in Figure~\ref{fig:Parity}c. We plot $\Pi(t_g)$ after the microwave $\pi/2$ pulse, and observe oscillations in the parity signal as the gradient time $t_g$ is scanned. The contrast of these oscillations is consistent with what is expected given the exchange oscillation contrast in Figure~\ref{fig:exchange}, and non-vanishing parity oscillation would certify entanglement in the ideal case of perfect spin preparation. However, we have imperfect spin preparation, and the erroneous spin populations outside the $\{\vert \!\!\downarrow \rangle_L \vert \!\!\uparrow \rangle_R,\vert \!\!\uparrow \rangle_L \vert \!\!\downarrow \rangle_R \}$ manifold could lead to parity oscillations even in the absence of entanglement. We have derived a condition on the parity oscillation contrast that is necessary and sufficient to certify entanglement and is the simplest way to see there is entanglement in our system (for a full derivation of this condition and its relation to other entanglement metrics see supplementary materials).  We relate the measured parity contrast, ${C}$, to the measured probabilities ($P^{\uparrow\uparrow}$,$P^{\downarrow\downarrow}$) that the spins are erroneously prepared in the same spin-state: If $C>C_{\rm bnd} = 4(P^{\uparrow\uparrow}P^{\downarrow\downarrow})^{1/2}$, then the state is entangled. By directly measuring the spin populations (see supplementary materials) and their associated uncertainty, we ascertain $C_{\rm bnd} = 0.133(25)$ as indicated by the dashed lines in Figure~\ref{fig:Parity}c. The observed parity oscillation contrast ${C}=0.49(4)$ exceeds $C_{\rm bnd}$ by more than $7\sigma$, certifying the presence of entanglement in the final state of the separated spins. We verify entanglement without correcting the measured parity for experimental imperfections, such as single atom loss due to background collisions.

While in Figure~\ref{fig:Parity}c we varied the parity detection parameters via $t_g$, in Figure~\ref{fig:Parity}d we measure the dependence of the parity on the exchange time at fixed $t_g$, thereby observing oscillations as the exchange interactions periodically entangle and unentangle the two atoms. We fix $t_g$ in the parity detection such that the entangled state $\vert\psi_+ \rangle$ (gray lines in Figure~\ref{fig:Parity}b,d) is rotated to a peak in $\Pi$, corresponding to the creation of the triplet (blue lines in Figure~\ref{fig:Parity}b,c). Because this $t_g$ amounts to a $\pi/2$ rotation about the $z$-axis of the Bloch sphere, it will also rotate $\vert \psi_- \rangle$ to the singlet, which corresponds to maximal negative parity.  In the lower panel of Figure~\ref{fig:Parity}d, we show how the parity measured under these conditions oscillates at the exchange frequency $J_{\rm ex}/(2\pi\hbar)$ . For comparison, in the upper panel, we show the measured exchange oscillations (purple, green) without the parity detection.  At the linear points of the exchange oscillations, one expects maximal entanglement corresponding to states $\vert\psi_+ \rangle$ (gray) and $\vert\psi_- \rangle$ (orange) and thus the extremal points of the parity.  At the minima and maxima of the exchange oscillations, the atoms are unentangled and the parity vanishes.

In conclusion, we have demonstrated the entanglement of remote qubits using spin-exchange interactions and a general protocol for detecting spin entanglement in a diversity of systems. In a large register, tuning the entanglement phase could be achieved by a far-detuned focused probe that changes the local effective magnetic field experienced by a single qubit, and the qubits can be arranged to allow the passage of the optical tweezers without perturbing the qubits~\cite{Weitenberg2011,Weitenberg2011a}. While in this work we focus on quantum information applications, our abilities to control spin and motion of individual neutral atoms will allow the study of intriguing microscopic models in condensed-matter physics, such as the Kondo Lattice model~\cite{Stewart1984}. %

  \textbf{Acknowledgements} This work was supported by the David and Lucile Packard Foundation and the National Science Foundation under grant number 1125844.  CAR acknowledges support from the Clare Boothe Luce Foundation. MLW and AMR acknowledge funding from NSF-PIF, ARO, ARO-DARPA-OLE, and AFOSR. MLW and MFF acknowledges support from the NRC postdoctoral fellowship program. 
 
 \textbf{Corresponding authors} Adam M. Kaufman or Cindy A. Regal: akaufman@physics.harvard.edu, regal@colorado.edu



\setcounter{figure}{0}
\setcounter{equation}{0}
\renewcommand{\theequation}{S\arabic{equation}}
\renewcommand{\thefigure}{S\arabic{figure}}
\renewcommand{\thetable}{S\arabic{table}}


\begin{center}
  \textbf{Supplementary Materials}
\end{center}
\section{State preparation and experimental protocol}

The experiment begins with atoms loaded stochastically from a magneto-optical trap into a pair of optical tweezers separated by $1.57~\mathrm{\mu m}$, each with an optical waist of $710~\mathrm{nm}$ and a $23(1)~\mathrm{MHz}$ depth~\cite{Schlosser2001}.  For all data presented in the figures, we post-select on experiments in which each optical tweezer is loaded with a single atom, based upon atom population (0 or 1) measurements in each tweezer at the start of the experiment.  We perform optical pumping and three-dimensional Raman sideband cooling as described in Refs.~\cite{Kaufman2012,Kaufman2014}, leaving each atom in the motional ground state of its respective tweezer with $0.90(7)$ fidelity, as determined via motional sideband spectroscopy, and in the $\ket{F,m_F} = \ket{2,2} \equiv \ket{\!\uparrow}$ spin state. To perform single-atom addressing, we subsequently apply circularly-polarized light to the left tweezer to induce a relative shift between the tweezers of the $\ket{\!\uparrow} \leftrightarrow \ket{1,1}\equiv\ket{\!\downarrow}$ transition~\cite{Weitenberg2011,Kaufman2014}. For the data in Figure~\ref{fig:exchange}b and Figure~\ref{fig:Parity}d, we apply a resonant square microwave $\pi$-pulse to the non-light-shifted atom, yielding a preparation fidelity in the desired two-particle spin configuration of 0.835(10). For the data in Figure~\ref{fig:Parity}c, we improve this fidelity by using temporally shaped Gaussian pulses in order to minimize $C_{\rm bnd}$ via the suppression of off-resonant transitions.

After motional and spin preparation, the position of the left optical tweezer is swept adiabatically in 10 ms to the other tweezer, realizing a two-atom spacing (Gaussian centroid separation) of $715~\mathrm{nm}$ ($854~\mathrm{nm}$). The tweezer depths are then reduced to $91(4)$ kHz in an adiabatic ramp~\cite{Kaufman2014}; in this configuration, there is a measured resonant tunneling of $J_{\rm eg}/(2\pi \hbar) = 165(6)~\mathrm{Hz}$ between the left-tweezer ground state and right-tweezer excited state. 

We then apply the first adiabatic passage (AP) through the tunneling resonance. Starting with the left optical tweezer tuned $2.2~\mathrm{kHz}$ below the tunneling resonance, the left tweezer depth is swept linearly in $12~\mathrm{ms}$ to symmetrically above the resonance. The end of this ramp marks the nominal $t=0$ of the exchange dynamics, and the resulting tweezer depths are static during the exchange time. After the exchange time, the AP is performed exactly in reverse, transferring the motional-excited atom back into its origin tweezer. The full AP procedure in both directions has a measured fidelity of $0.81(4)$ for the data in Figure~\ref{fig:exchange}b; we observe systematic fluctuations in this fidelity such that its value is $0.69(2)$  for the data in Figure 3c.

At this point, if we are directly measuring the exchange dynamics as in Figure~\ref{fig:exchange}, the tweezers are ramped up in depth and separated for spin detection and imaging. If we are performing entanglement detection, instead directly we apply the magnetic-field gradient.  After switching on the gradient there is a 3 ms delay for field settling.  Then we hold the gradient on for variable time $t_g$, after which the gradient is switched off.  After another 3 ms delay to let the fields settle, we apply the microwave $\pi/2$ pulse for parity read-out, after which the tweezers are increased in depth and separated for spin detection and imaging.

The observed spin-exchange contrast can be influenced by a number of factors, including the single-particle ground-state fraction, single-particle loss, and the AP fidelities. Spin preparation also affects the contrast, but as discussed in the text the data presented in Figure~\ref{fig:exchange}b are post-selected on the desired anti-aligned spin configuration (for the parity measurements, post-selection is not possible since we study all final spin configurations).  Based on our measurements of experimental systematics -- the ARP fidelity of 0.81(4), the single-particle ground-state fraction of 0.90(7), and the single-particle survival probability of 0.963(7) -- and combined with the spin-preparation post-selection, we would expect a spin-exchange contrast of $0.60(10)$, which exceeds the measured value of $0.29(2)$ in Figure~\ref{fig:exchange}b. In the absence of the spin-preparation post-selection, we observe a spin-exchange contrast that is reduced by the spin-preparation fidelity of 0.835(10), as expected. 

For the characterization of the spin-populations, in a distinct experiment we measure the spin-down probability of the atom in each tweezer by applying resonant push-out light directly prior to when the first AP of the experiment would occur. Accounting for the separately measured single-atom survival probability of 0.963(7), we measure $\{P^{\uparrow\uparrow}$,$P^{\downarrow\downarrow} \} =\{0.071(14), 0.016(5)\}$, leading to $P^{\uparrow\downarrow}+P^{\uparrow\downarrow} = 1-(P^{\uparrow\uparrow}+P^{\downarrow\downarrow})$ . These numbers are used to calculate $C_{\rm bnd}$ and $F$. The measured parity data is not corrected for loss of single atoms due to background collisions, which degrades the entanglement generation.

\section{Modeling of interactions and tunneling during state preparation}

For two atoms of mass $m$ in different motional wave functions  $\psi_e ({\bf r}) = \langle {\bf r} \vert e\rangle$ and $\psi_g ({\bf r}) = \langle {\bf r} \vert g\rangle$,  the exchange interaction energy $J_{ex}=2 U_{eg}$, where $U_{eg}=\frac{4\pi\hbar^2 a_s}{m} \int \vert \psi_e({\bf r})\vert ^2 \vert \psi_g({\bf r})\vert^2 d^3{\bf r}$  is the contact interaction energy proportional to $a_s$, the atomic $s$-wave scattering length.  To compute this integral we use efficient numerical techniques to determine the spectrum and eigenstates of the 3D optical tweezer potential~\cite{Wall2015} using parameters from an independent experimental characterization~\cite{Kaufman2014}.  This calculation, combined with experimental uncertainties, gives rise to the blue theoretical swath in Figure~\ref{fig:exchange}.

\begin{figure}[!h]
	\centering
	\includegraphics[scale=0.5]{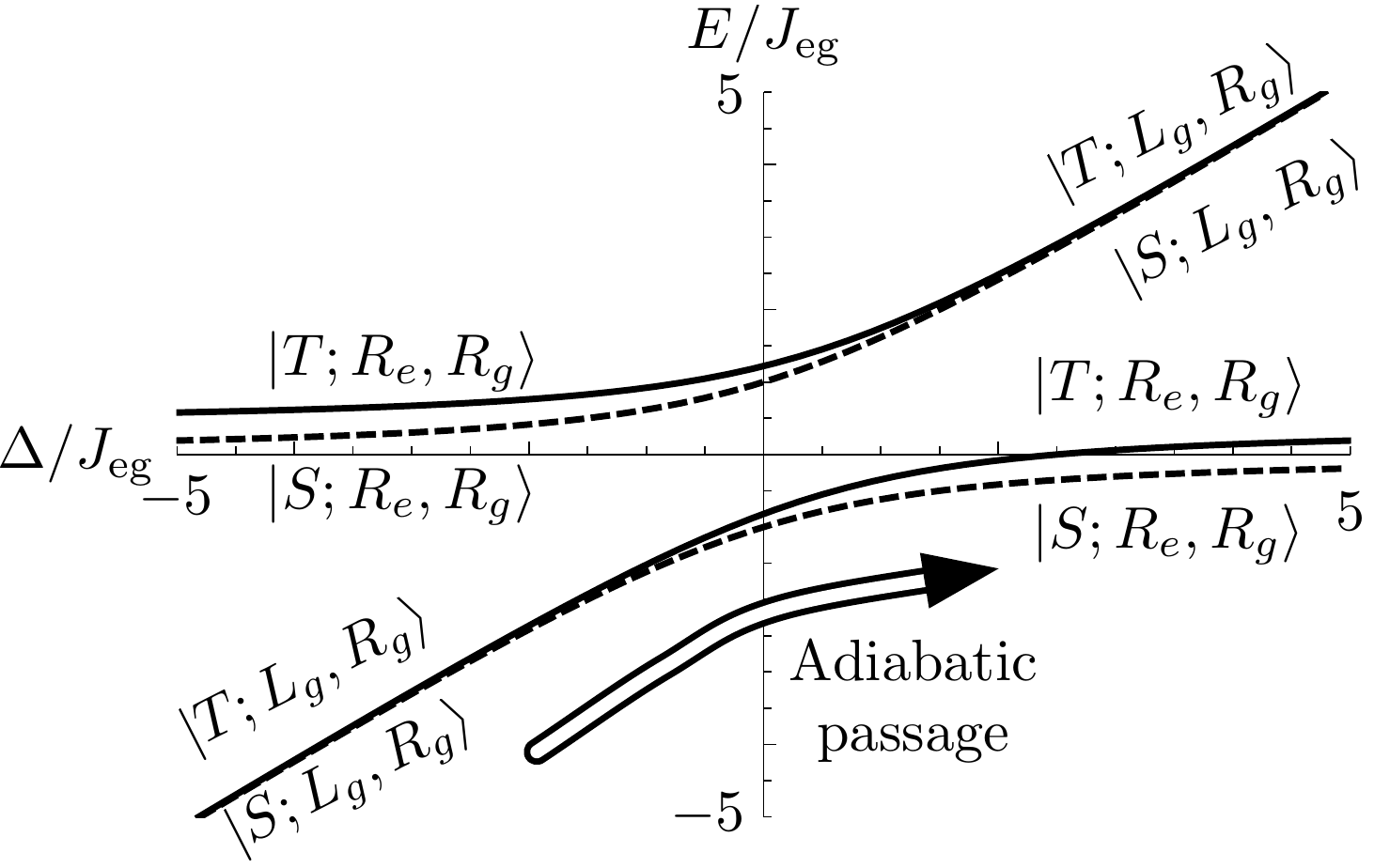}
	\caption{Adiabatic energy-eigenstates as a function of the double-well bias $\Delta$ in units of the ground-excited tunneling $J_{\rm eg}$. At large positive bias, the triplet and singlet eigenstates corresponding to two particles in the same well are split by $J_{ex}$.}
	\label{fig:ARP}
\end{figure}

To theoretically model the adiabatic passage procedure (AP), we consider a situation in which two particles with opposite spins are initially localized in the ground motional states of the left and right wells, denoted $L_g$ and $R_g$, respectively.  During the AP, the bias between wells $\Delta$ is tuned so that $L_g$ is near resonance with an excited level of the right well, denoted $R_e$.  In what follows, we will measure the bias $\Delta$ with respect to this resonance position, i.e.~the resonance occurs at $\Delta=0$.  Further, we consider that the bias range is such that tunnel-couplings to all other motional states (e.g., $L_g\to R_g$ tunneling) are negligible, and so we can restrict ourselves to the set of motional states spanned by $L_g$, $R_g$, and $R_e$ and the single $L_g\to R_e$ tunneling resonance at $\Delta=0$.  Because of the mixing of spin components due to the spin exchange interaction, it is most convenient to use the singlet-triplet basis $\{|S; L_g,R_g\rangle, |S;R_e,R_g\rangle ,|T; L_g,R_g\rangle, |T;R_e,R_g\rangle \}$, where $S$ and $T$ denote singlet and triplet states and the latter two labels are the motional states of the two particles.  In this basis, the Hamiltonian is
\begin{align}
\hat{H}&=\left(\begin{array}{cccc} \Delta&-J_{\rm eg}&0&0\\ -J_{\rm eg}&0&0&0\\ 0&0&\Delta &-J_{\rm eg}\\ 0&0&-J_{\rm eg}&2U_{eg}\end{array}\right)\, ,
\end{align}
where $J_{\rm eg}$ is the tunneling amplitude for the process $L_g\to R_e$, and the initial state at large negative $\Delta$, $|L_g\uparrow,R_g\downarrow\rangle$, is an equal weight superposition of $|S; L_g,R_g\rangle$ and $|T; L_g,R_g\rangle$.  The AP process is described by separate tunneling avoided crossings in the singlet and triplet channels, as shown in Figure~\ref{fig:ARP}.  Provided that the ramping procedure is adiabatic with respect to these avoided crossings, it will transfer the initial state into an equal weight superposition of $|S; R_e,R_g\rangle$ and $|T; R_e,R_g\rangle$ at large positive $\Delta$.  This pair of eigenstates, which correspond to the two particles occupying the same well, have an asymptotic energy splitting of $2U_{eg}$ as $|\Delta|\gg J_{\rm eg}$.  Note that the position of the bias resonance for the triplet channel is shifted by $2U_{eg}$ with respect to the resonance in the singlet channel.  The energy offset does not affect the degree of adiabaticity of the ramping procedure  and just gives rise to a phase shift between the singlet and triplet components.  

\section{Entanglement verification based on parity oscillations}

Here we derive a criterion for verifying entanglement generated by spin-exchange interactions in a two-atom system.  Our strategy will be to assume an unentangled (separable) density matrix, and from this assumption establish a constraint on experimentally measurable quantities: the parity oscillation contrast (Fig.\ 3c) and the populations of different spin states.  Experimental violation of this constraint thus verifies entanglement.

In the experiment, spin-exchange occurs between two atoms occupying a single optical tweezer, and then those atoms are separated into two tweezers.  Because the adiabatic passages in the experiment are imperfect, the atoms may sometimes end up in the same tweezer after the attempted separation.  For clarity of presentation, we first consider the idealized case of perfect adiabatic-passage fidelity; hence $\rho$ in what follows describes states immediately after the second adiabatic passage in which one atom occupies each tweezer.  At the end of this section we carefully consider the effects of adiabatic passage failure, and show that they do not affect our claims of entanglement.

Because the measured spin-coherence time in the experiment is much less than the time between when we prepare the initial spin state and when we complete the adiabatic passages, $\rho$ cannot have any coherences between states with different total $S^z=S^z_L+S^z_R$.  Working in a basis that diagonalizes both $\hat{S}_{L}^z$ and $\hat{S}_{R}^z$, $\{|\!\uparrow\rangle_{L}|\!\uparrow\rangle_{R},|\!\uparrow\rangle_{L}|\!\downarrow\rangle_{R},|\!\downarrow\rangle_{L}|\!\uparrow\rangle_{R},|\!\downarrow\rangle_{L}|\!\downarrow\rangle_{R}\}$, the most general density matrix satisfying this condition can be written
\begin{align}
\label{eq:rho}
\rho=\left(\begin{array}{cccc}
P(\uparrow_{L},\uparrow_{R}) & 0 & 0 & 0 \\
0 & P(\uparrow_{L},\downarrow_{R}) & \varepsilon & 0 \\
0 & \varepsilon^{*} & P(\downarrow_{L},\uparrow_{R})  & 0 \\
0 & 0 & 0 &P(\downarrow_{L},\downarrow_{R})
\end{array}\right).
\end{align}
The populations $P(\uparrow_{L},\uparrow_{R})$ and $P(\downarrow_{L},\downarrow_{R})$ are, respectively, the total probabilities of having both atoms in the $|\!\uparrow\rangle$ state or both atoms in the $|\!\downarrow\rangle$ state, and are non-zero due to imperfect initial spin preparation.  Because these probabilities are conserved by the adiabatic passages and the spin-exchange, their measured values \emph{before} the adiabatic passage, referred to as $P^{\uparrow\uparrow}$ and $P^{\downarrow\downarrow}$ in the manuscript, can safely be used in Eq.\ (\ref{eq:rho}): $P(\uparrow_{L},\uparrow_{R})=P^{\uparrow\uparrow}$ and $P(\downarrow_{L},\downarrow_{R})=P^{\downarrow\downarrow}$.  The parity is measured after first applying a magnetic-field gradient for a variable time $t_g$, and then applying a $\pi/2$ microwave pulse, which transforms $\rho\rightarrow \tilde{\rho}(t_g)$.  After some algebra, the parity of $\tilde{\rho}(t_g)$ can be written $\Pi(t_{\rm g})=2{\rm Re}(e^{-i\delta t_{\rm g}}\varepsilon)$, which oscillates as a function of $t_{\rm g}$ with a contrast of $C=4|\varepsilon|$.  

Our goal is to derive a constraint on $C$ in terms of the measured quantities $P^{\uparrow\uparrow}$, $P^{\downarrow\downarrow}$, under the assumption that $\rho$ is separable.  If $\rho$ were a product state $\rho_{\rm L}\otimes\rho_{\rm R}$, where
\begin{align}
\rho_{L(R)}=\left(\begin{array}{cc}
\rho^{\uparrow,\uparrow}_{L(R)}  & \rho^{\uparrow,\downarrow}_{L(R)} \\
\rho^{\downarrow,\uparrow}_{L(R)} & \rho^{\downarrow,\downarrow}_{L(R)} \\
\end{array}\right),
\end{align}
then
\begin{align}
\label{eq:ps_bound}
&|\varepsilon|=|\rho_{L}^{\uparrow,\downarrow}|\times|\rho_{R}^{\downarrow,\uparrow}|\leq \big(\rho_{L}^{\uparrow,\uparrow}\rho_{L}^{\downarrow,\downarrow}\big)^{1/2}\times\big(\rho_{R}^{\uparrow,\uparrow}\rho_{R}^{\downarrow,\downarrow}\big)^{1/2}\\
&=\big(P(\uparrow_{L},\uparrow_{R})P(\downarrow_{L},\downarrow_{R})\big)^{1/2}=\big(P^{\uparrow\uparrow}P^{\downarrow\downarrow}\big)^{1/2}.
\end{align}
By definition, a separable state can be written as a classical mixture of product states, $\rho=\sum_{j}\lambda^j\rho_{L}^j\otimes\rho_{R}^j$, in which case
 \begin{align}
&|\varepsilon|\leq\sum_{j}\lambda_j|\varepsilon_j|\leq\sum_{j}\lambda^j\big(P^{\uparrow\uparrow}_jP^{\downarrow\downarrow}_j\big)^{1/2}\\
&\leq \big(\sum_j\lambda^j P^{\uparrow\uparrow}_j\big)^{1/2}\big(\sum_j\lambda^j P^{\downarrow\downarrow}_j\big)^{1/2}=\big(P^{\uparrow\uparrow}P^{\downarrow\downarrow}\big)^{1/2}.
\end{align}
Here, the second inequality is the constraint on $|\varepsilon|$ derived for a product state in Eq.\ (\ref{eq:ps_bound}) applied to each state in the classical mixture, and the first and third are triangle inequalities.  Using $C=4|\varepsilon|$, we therefore have guaranteed entanglement whenever
\begin{equation}
\label{eq:bound}
C>4\big(P^{\uparrow\uparrow}P^{\downarrow\downarrow}\big)^{1/2}.
\end{equation}

We can also verify entanglement by using an entanglement witness associated with the fidelity of $\rho$ in the maximally entangled state $|\Psi_{+}\rangle=(|\!\uparrow\rangle_{L}|\!\downarrow\rangle_{R}+i|\!\downarrow\rangle_{L}|\!\uparrow\rangle_{R})/\sqrt{2}$, $F=\langle\Psi_+|\rho|\Psi_+\rangle$.  As shown in Ref.\ \cite{Sackett2000}, $\rho$ is entangled if $F>1/2$.  In terms of Eq.\ (\ref{eq:rho}), the fidelity can be written
\begin{align}
F&=\frac{1}{2}[P(\uparrow_{L},\downarrow_{R})+P(\downarrow_{L},\uparrow_{R})]+\frac{C}{4}\\
&=\frac{1}{2}+\frac{C}{4}-\frac{1}{2}(P^{\uparrow\uparrow}+P^{\downarrow\downarrow}).
\end{align}
Therefore $F>1/2$ is equivalent to $C> 2(P^{\uparrow\uparrow}+P^{\downarrow\downarrow})$, which agrees with Eq.\ (\ref{eq:bound}) when $P^{\uparrow\uparrow}=P^{\downarrow\downarrow}$.  Equation (\ref{eq:bound}) can also be derived by applying the Peres-Horodecki criterion to $\rho$,  and thus it is actually both sufficient \emph{and} necessary for entanglement~[33, 34].  Thus, in contrast with the fidelity-based entanglement witness $C> 2(P^{\uparrow\uparrow}+P^{\downarrow\downarrow})$, the right-hand-side of Eq.\ (\ref{eq:bound}) is as small as possible for any $P^{\uparrow\uparrow}$ and $P^{\downarrow\downarrow}$, resulting in the greatest possible confidence in entanglement for a particular measured contrast.  We also note that the extent to which Eq.\,(\ref{eq:bound}) is satisfied, $\frac{1}{2}\big(C-4\big(P^{\uparrow\uparrow}P^{\downarrow\downarrow}\big)^{1/2}\big)>0$, is a direct measurement of the concurrence in the density matrix $\rho$ \cite{Endres}.

\subsection{Imperfect adiabatic passages}

As mentioned above, the adiabatic passages in the experiment are not perfect.  For our purposes, we define success as any situation in which the atoms end up in different tweezers, which occurs either if both of the individual adiabatic passages succeed or if they both fail.  Conversely, we define failure as any situation in which both atoms end up in the same tweezer, which happens if one of the adiabatic passages is successful while the other is not.  While the states resulting from failure have not been considered in deriving Eq.\ (\ref{eq:bound}), they do not contribute to the parity oscillation contrast, since two atoms in the same tweezer are not sensitive to a magnetic-field gradient.  Therefore, we intuitively expect that imperfect adiabatic passage can only \emph{lower} the measured contrast, such that Eq.\ (\ref{eq:bound}) still implies entanglement.

This intuition can be formalized by introducing a projector, $\hat{K}$, onto the states with one atom in each tweezer.  Defining $f$ as the success probability, we can then form projections of the true experimental density matrix, denoted by $\rho_{\rm exp}$, into the subspaces defined by success or failure:
\begin{align}
\label{eq:projections}
\hat{K} \rho_{\rm exp} \hat{K}&\equiv f\rho_{\rm succ}\\
(\mathds{1}-\hat{K})\rho_{\rm exp}(\mathds{1}-\hat{K})&\equiv(1-f)\rho_{\rm fail}.
\end{align}
Note that, by the choice of pre-factors, both $\rho_{\rm succ}$ and $\rho_{\rm fail}$ are properly normalized density matrices.  Importantly, $\rho_{\rm succ}$ has precisely the form given in Eq.\ (\ref{eq:rho}); we can therefore directly apply the arguments above, deducing that $\rho_{\rm succ}$ is entangled whenever $|\varepsilon|>\sqrt{P^{\uparrow\uparrow}P^{\downarrow\downarrow}}$.  The only difference from before is in the relationship between $|\varepsilon|$ and the measured parity oscillation contrast.  The application of a magnetic-field gradient and the subsequent $\pi/2$ pulse does not mix the two subspaces partitioned by $\hat{K}$, and therefore the measured parity can be written $\Pi(t_g)=f\Pi_{\rm succ}(t_g)+(1-f)\Pi_{\rm fail}$.  Importantly, $\Pi_{\rm fail}$ is \emph{independent} of $t_g$ because two atoms in the same tweezer are not sensitive to a magnetic-field gradient, and therefore the measured parity oscillation contrast is given simply by $C=f 4|\varepsilon|$.  Thus the effect of imperfect adiabatic passages is that, in terms of the measured contrast, the condition $|\varepsilon|>\sqrt{P^{\uparrow\uparrow}P^{\downarrow\downarrow}}$ now reads $C/f>4\sqrt{P^{\uparrow\uparrow}P^{\downarrow\downarrow}}$  [c.f. Eq.\ (\ref{eq:bound})].  The criterion used in the manuscript, $C>4\sqrt{P^{\uparrow\uparrow}P^{\downarrow\downarrow}}$, is therefore a conservative way to verify entanglement in $\rho_{\rm succ}$, since $C/f$ is strictly larger than $C$.

The fidelity can now be written $F=\langle\Psi_+|\rho_{\rm exp}|\Psi_+\rangle=f\langle\Psi_+|\rho_{\rm succ}|\Psi_+\rangle\equiv f F_{\rm succ}$, where $F_{\rm succ}$ is the fidelity of the (projected and normalized) density matrix $\rho_{\rm succ}$, and can be equivalently written as $F_{\rm succ}=\frac{1}{2}+|\varepsilon|-\frac{1}{2}(P^{\uparrow\uparrow}+P^{\downarrow\downarrow})$.  Similar to before, entanglement of $\rho_{\rm succ}$ is now guaranteed by $F_{\rm succ}>1/2$, or equivalently $4|\varepsilon|> 2(P^{\uparrow\uparrow}+P^{\downarrow\downarrow})$.  Using $4|\varepsilon|=C/f$, we can extract $F_{\rm succ}$ from the measured contrast [$C=0.49(4)$] and measured success probability [$f=0.69(2)$], obtaining $F_{\rm succ}=0.634(17)$.  Note that the \emph{actual} fidelity $F$ is reduced from this value by the success probability $f$, and is in fact below $1/2$.  However, this is not inconsistent with entanglement in $\rho_{\rm exp}$, as the fidelity-based witness for $\rho_{\rm exp}$, written in terms of the actual fidelity, is $F>f/2$.



 \hspace{1cm}

\noindent [33]~~A. Peres, {\em Separability Criterion for Density Matrices}, Phys. Rev. Lett.
  {\bf 77},  1413  (1996).

\noindent [34]~~M. Horodecki, P. Horodecki, and R. Horodecki, {\em Separability of Mixed
  States: Necessary and Sufficient Conditions}, arXiv:quant-ph/9605038  (1996).

\end{document}